\documentclass{aastex61}
\listofchanges

\newcommand\aastex{AAS\TeX}

\submitjournal{ApJ}

\shorttitle{\aastex\ Nova SMC 2016-10a}
\shortauthors{Orio et al.}

\begin{document}

\title{What we learn from the X-ray grating spectra of Nova SMC 2016
}

\correspondingauthor{Marina Orio}
\email{orio@astro.wisc.edu}

\author{M. Orio}
\affiliation{Department of Astronomy, University of Wisconsin,
475 N. Charter Str., Madison WI 53706, USA}
\affiliation{INAF-Osservatorio Astronomico di Padova, vicolo Osservatorio, 5,
 35122 Padova, Italy}
\author{J.-U. Ness}
\affiliation{XMM-Newton Science Operations Center, European Space Astronomy Center, Camino Bajo del Castillo s/n, Urb. Villafranca del Castillo, Villanueva de la Cañada, E-28692 Madrid, Spain}
\author{A. Dobrotka}
\affiliation{Advanced Technologies Research Institute, Faculty of Materials Science and Technology in Trnava, Slovak University of Technology in Bratislava, Bottova 25, 917 24 Trnava, Slovakia}
\author{E. Gatuzz}
\affiliation{ESO, Karl-Schwarzschild-Strasse 2, D-85748 Garching bei M\"unchen, Germany}
\author{N. Ospina }
\affiliation{via Pietro Pomponazzi, 33, 35124 Padova, Italy}
\author{E. Aydi}
\affiliation{South African Astronomical Observatory, P.O. Box 9, 7935 Observatory, South Africa}
\author{E. Behar}
\affiliation{Department of Physics, Technion, Haifa, Israel}
\affiliation{Astronomy Department, University of Cape Town, 7701 Rondebosch, South Africa}
\author{D.A.H. Buckley}
\affiliation{South African Astronomical Observatory, P.O. Box 9, 7935 Observatory, South Africa}
\author{S. Ciroi}
\affiliation{Dipartimento di Fisica e Astronomia, vicolo Osservatorio, 3, 35122 Padova, Italy}
\author{M. Della Valle}
\affiliation{INAF-Osservatorio Astronomico di Capodimonte, Salita Moiarello, 16, I-80131
Napoli, Italy} 
\author{M. Hernanz}
\affiliation{Institut de Ci\`encies de l' Espai (ICE-CSIC). Campus UAB. c/ Can Magrans s/n, 08193, Bellaterra, Spain}
\affiliation{Institut d'Estudis Espacials de Catalunya, c/Gran Capit\`a 2-4, Ed. Nexus-201, 08034, Barcelona, Spain}
\author{M. Henze}
\affiliation{Department of Astronomy, San Diego State University, San Diego, CA 92182, USA}
\author{J.P. Osborne}
\affiliation{X-ray and Observational Astronomy Group, Department of Physics \& Astronomy, University of Leicester, LE1 7RH, UK}
\author{K.L. Page}
\affiliation{X-ray and Observational Astronomy Group, Department of Physics \& Astronomy, University of Leicester, LE1 7RH, UK}
\author{T. Rauch}
\affiliation{Institute for Astronomy and Astrophysics,
 Kepler Center for Astro and Particle Physics. Eberhard Karls
 University, Sand. 1., 72076 T\"ubingen, Germany}
\author{G. Sala}
\affiliation{Departament de F`'isica, EEBE, Universitat Polit\`ecnica de Catalunya. BarcelonaTech.,
 Av. d' Eduard Maristany 10-14, 08019, Barcelona, Spain}
\affiliation{Institut d' Estudis Espacials de Catalunya, c/Gran Capit\`a 2-4, Ed. Nexus-201, 08034, Barcelona, Spain}
\author{S. Starrfield}
\affiliation{School of Earth and Space Exploration, Arizona State University, Tempe, AZ 85287-1404, USA}
\author{R.E. Williams}
\affiliation{Space Telescope Science Institute, 3700 San Martin Drive, Baltimore, MD 21218 USA} 
\author{C.E. Woodward}
\affiliation{Minnesota Institute for Astrophysics, University of Minnesota, 116 Church Street, SE, Minneapolis, MN 55455, USA}
\author{P. Zemko}
\affiliation{via dei Tadi, 5, 35139 Padova, Italy}


\begin{abstract}
 Nova SMC 2016 has been the most luminous nova known in the direction
 of the Magellanic Clouds.  It turned into a very luminous supersoft X-ray
 source between day 16 and 28 after the optical maximum. We observed 
 it with {\sl Chandra}, the HRC-S camera and the Low Energy Transmission Grating
 (LETG) on 2016 November and 2017 January (days 39 and 88 after 
 optical maximum), and with {\sl XMM-Newton} on
 2016 December (day 75). We detected the compact white dwarf 
 (WD) spectrum as a luminous supersoft X-ray continuum with deep absorption features
 of carbon, nitrogen, magnesium, calcium, probably argon and sulfur on day 
 39, and  oxygen, nitrogen and carbon on days 75 and 88. The
spectral features attributed to the WD atmosphere are all blue-shifted, by about 1800
 km s$^{-1}$ on day 39 and up
 to 2100  km s$^{-1}$ in the following observations.
 Spectral lines attributed to low ionization potential
 transitions in the interstellar medium are also observed.
 Assuming the distance of the Small Magellanic
 Cloud, the bolometric luminosity exceeded Eddington level for at least three months.
 A preliminary analysis with atmospheric models indicates effective
 temperature around 700,000 K on day 39, peaking at the later dates 
 in the 850,000-900,000 K range, as
expected for a $\simeq$1.25 M$_\odot$ WD.
 We suggest a possible classification as an oxygen-neon WD,
 but more precise modeling is needed to accurately determine the abundances. 
 The X-ray light curves show large, aperiodic flux variability, 
 not associated with spectral variability.
 We detected red noise, but did not find periodic or quasi-periodic modulations.
\end{abstract}
\section{Introduction} \label{sec:intro}
Novae in outburst  are among the most luminous stellar X-ray sources
 in the sky. A recent overview of the observational
 facts has been given by \citet{Poggiani2018}.  
 Comprehensive reviews of the models can be found in \citet{starrfield2016,
 starrfield2012, prialnik2005}, and the basic facts can
 be summarized as follows. 
 Nova eruptions are due to thermonuclear burning of hydrogen via the
 CNO cycle, at the bottom of  a shell accreted by a white
 dwarf (WD) from a close binary companion. 
The outburst is repeated after quiescent periods
 ranging from few years to $\simeq$hundreds of thousands years.
The burning is ignited in conditions of electron degeneracy, and
 becomes explosive, inflating and possibly immediately ejecting
 part of the envelope.    Since the initial
 suggestion of \citet{Bathandshaviv}, many authors working
 on models have predicted that 
  the bulk of the remaining envelope mass  is then stripped by a
radiation-pressure driven wind,  although
  a wind may also be triggered by Roche Lobe overflow
\citep[see, for a discussion,][]{Wolf2013}.
Then, the evolutionary track of a post-nova is driven by a shift in the
 wavelength of maximum energy towards shorter wavelengths,
 at constant bolometric luminosity close to 10$^{38}$ erg s$^{-1}$
 \citep[e.g.][]{starrfield2012, Wolf2013}.  This phase  lasts from one week to $\approx$10 years
 as the WD photosphere shrinks close to pre-outburst radius, while thermonuclear
 burning is still occurring near the surface.
 Thus, post-outburst novae offer {\it a unique possibility to
observe the effects of nuclear burning near the stellar surface}.
 The WD reaches an effective temperature T$_{\rm eff}>$200,000 K,
  emitting in the X-ray range \citep{starrfield2012, Wolf2013} and becomes a supersoft X-ray
 source (SSS).  Despite its large luminosity, the X-ray
 flux of the burning WD is very easily absorbed. If the column
 density exceeds 10$^{22}$
 cm$^{-2}$, it may never be detected. In most cases,
 however, the  column density and/or
 the filling factor of the ejecta  must be sufficiently low
 that the WD becomes observable in X-rays at large
 distances, at the outskirts of the Local Group and beyond,
 wherever the interstellar column density is low \citep[see][and
 references therein]{orio2012, henze2014}.

 There is a second source of X-ray flux in novae in outburst, with different
 time scale and evolution in each nova: it is the  X-ray emission that originates early in the
 outburst from the ejecta, most
 likely because of violent shocks in colliding winds \citep[][and references
 therein]{orio2012}.
 In most novae, the peak luminosity of this emission from the shell is
 10$^{34}$ erg s$^{-1}$ \citep[see e.g.][]{orio1996, orio2001,
 peretz2016}. In symbiotic novae, in which the
 secondary is a red giant,
the impact of the ejecta with the giant wind produces thermal X-ray emission
 with a peak luminosity even as high as 10$^{36}$ erg s$^{-1}$ 
\citep{nelson2008, ness2007}.
The initial temperature of the ejecta exceeds 10 keV, but 
the shocked plasma is observed to cool with time until the emission
 lines are only in the very soft range. Often the soft X-rays' emission
 lines are superimposed on the spectrum of the central source \citep[see][and references therein]{orio2012, orio2013}.  

 Novae in the Magellanic Clouds are sufficiently close to us
 and in directions of such low column density that we can
 obtain X-ray grating spectra to study them in detail, and offer us the opportunity 
 to study targets at known distance and compare the population
 of Galactic novae with one in a much lower metallicity environment.

\section{An intriguingly ``hyper-luminous'' nova}
A comprehensive description of the observational facts concerning N SMC 2016 
 can be found in \citet{aydi2018}, so we will give here only a brief summary.
The nova was discovered by MASTER (as OT  J010603.18- 744715.8)  on
  2016 October 14 \citep{Shumkov2016}, but the outburst had started earlier,
  with maximum
 magnitude detected on 2016 October 9 \citep{Jablonski2016, Lipunov2016}. The latter is the
 date from which we count the post-outburst days in our observations' list in  Table 1.
N SMC 2016 was a very fast nova according to the classification
 by \citet{Warner2003}, with  the  time
 for a decline by two visual magnitudes, t$_2$=4$\pm$1.0 d, a full width
 at half maximum
 (FWHM) of the Balmer lines of 3500$\pm$100 km s$^{-1}$, and an amplitude of 12.1$\pm$1 visual
 magnitudes.

The most striking characteristic of this nova is its extreme peak luminosity,
 larger than any known nova in the Magellanic Clouds. In the 
 MASTER Very Wide Field Camera, which measures visual magnitude
 almost exactly corresponding to Johnson V (unless the H$\alpha$
 line was already prominent, skewing the spectral energy
 distribution, but this would be unusual 
 already at maximum), the nova
 reached V=8.9$\pm$0.3 on HJD 2457671.3 on 2016 October 9
 (note that the value given by \citet{Lipunov2016} has been corrected by
 Lipunov, in a private communication to us).  If we assume an
average distance modulus 18.96 \citep[][corresponding to 62 kpc]{Scowcroft2016},
 and only moderate reddening, we find an absolute magnitude 
 V$\simeq$-10.1, about 11 times the Eddington luminosity for a 1 M$_\odot$ star.
 The optical and UV light curves presented in \citet{aydi2018}
 imply super-Eddington luminosity for at least 88 days,  
 but the {\sl SWIFT} X-Ray Telescope (XRT) light curve 
 analyzed in that article, that we show
  again here in Fig. 1  to illustrate at what stages
 our observations were obtained, indicates  that in the SSS stage
 the X-ray luminosity (representative of most of the bolometric luminosity), 
 did not decline significantly for about 6 months. 
 As discussed by \citet{aydi2018} the SMC has a certain extension along the line of sight, so the distance
 modulus could be as low as 18.7 if the nova is on the side closer to us,
 but the luminosity would still be largely super-Eddington. 
 In order to check the distance to verify
 whether the prolonged super-Eddington period was real,
 \citet{aydi2018} evaluated instead the maximum absolute magnitude with the
 maximum magnitude versus rate of decline  (MMRD) method and other 
 empiric formulations, and concluded that the resulting distance would be only
 d$\simeq 42^{+8}_{-7}$ kpc, which would put the nova in front of the galaxy.
 However, this seems unlikely, so \citet{aydi2018} still favored SMC
 distance.  \citet{DellaValle1991} found already
 back in 1991 that the few very intrinsically luminous and fast novae,
 like N SMC 2016,
 do not follow the MMRD, so this may explain the non-SMC distance obtained
 when we try and apply the MMRD to Nova SMC 2016.

  In fact, assuming that this nova                
 is in the SMC, very few novae have been as intrinsically
 luminous at maximum as N SMC 2016.  
 The Galactic nova CP Pup  had comparable peak absolute magnitude; 
 another Galactic nova, V1500 Cyg
  \citep[see][and references therein]{Shafter2009, Strope2010},
and at least
two extragalactic novae (M31N2007-11d and LMC 1991, see \citet{Schwarz2001}) 
 had maximum absolute magnitude V$\leq$-9. 
 The parallax has been measured precisely with Gaia for CP Pup 
 \citep[for the conversion of parallax to distance, see][]{bailerjones2018,
Luri2018}, and its 
 large luminosity, around maximum absolute
magnitude, around V=-10, has been  confirmed.
 There seems to be also a distinct small population
 of superluminous novae in M31 
 of which the majority show Fe II type spectra and are fast,
 with t$_2<10$ days (Williams, presentation at EWASS2018 conference). 
  N SMC 2016 is thus likely to
 belong to a rare class of super-luminous novae, observed in
 the Galaxy and Magellanic Clouds only 
 once in $\simeq$25 years. 
 
 \citet{aydi2018} discussed how the derived parameters point at
 a massive WD, about 1.25 M$_\odot$ when compared with theoretical models
 by \citet[][]{Wolf2013, Hillman2016}.
\begin{table*}
\caption{Chandra and XMM-Newton observations of Nova SMC 2016 examined in
 this article, and measured count rates for the X-ray detectors. 
}
\label{table:obs}
\begin{center}
\begin{tabular}{rrrrrrr}\hline\hline \noalign{\smallskip}
 Instrument & ObsID & Exp. time$^a$ & Date$^b$ &  Day$^c$ & c.r.         & L$_{\rm X}$ (60kpc)$^d$ \\
            &       & (ks)          &          &          & cts s$^{-1}$ &   erg s$^{-1}$ \\
\hline \noalign{\smallskip}
 Chandra HRC-S+LETG & 19011     & 30.16 & 2016-11-17 & 39  & 6.589$\pm$0.011  & 3.42 $\times 10^{38}$ \\
 XMM-Newton RGS     & 079418020 & 29.80 & 2016-12-22 & 75  & 24.56$\pm$0.02   & 3.9 $\times 10^{38}$ \\
 Chandra HRC-S+LETG & 19012     & 28.16 & 2017-01-04 & 88  & 10.110$\pm$0.014 & 6.09  $\times 10^{38}$ \\ 
\hline \noalign{\smallskip}
\end{tabular}
\end{center}
Notes:\hspace{0.1cm} $^a $: Exposure time of the observation (dead time corrected);
 $^b $: Start date of the observation; $^c $: Time in days after the discovery on 2016-10-14;
$^d$: In the 10-60 \AA \ range for Chandra, in the 10-38 \AA \ range for the XMM-Newton RGS. \\
\end{table*}
\section{The X-ray evolution}
 Fig. 1 shows the X-ray light curve measured with the {\sl Swift} X-ray Telescope
 \citep[XRT:][]{Burrows2005}, 
 already presented by \citet{aydi2018}. The nova was not detected
 in the first two weeks after the outburst, on 2016 October 26 it became
 unobservable with the XRT for 11 days for the pole constraint 
 (which means that Swift cannot point too close to the Earth limb),
 then on 2016 November 7 a luminous
 supersoft X-ray source was detected.  Dense monitoring 
 followed, initially three times per day, then twice a day, with another
 period of non-observability in the last two weeks of December. The 
 X-ray source was variable, but 
 the average X-ray luminosity was clearly increasing until about day 40,  
 then almost a plateau followed with an apparent
 peak around day 83, when the luminosity started
 decreasing very slowly. A rapid decay started only around day 180 after the outburst.
 As Table 1 shows, a high resolution X-ray spectrum was first taken on 
 2016 November 17 (day 39), shortly after the SSS discovery (with {\sl Chandra},
 using the
 HRC-S camera and the Low Energy Transmission  Grating or LETG),
 and  few days before the maximum recorded count rate.
 It was observed again a few days after the maximum X-ray count rate 
on 2016 December 22 with ({\sl XMM-Newton}, day 75),
 and  on 2017 January 4 with the Chandra HRC-S+LETG, day 88).
The {\sl Chandra} HRC-S+LETG wavelength range is 1.2-175 \AA \ (corresponding
 to the energy range 0.07-10 keV), with a resolving power
 $\approx 20 \times \lambda$ in the
 3-50 \AA \ range (and $\lambda$/$\Delta\lambda>$1000 at 50-160 \AA),
effective area peaking at 25 cm$^{-2}$, and a resolution
 of 0.05 \AA. The {\sl XMM-Newton} RGS1 and
 RGS2 gratings probe the 5-38 \AA \ spectral range (corresponding to 
 0.35-2.5 keV) with maximum
 effective area of 61 cm$^2$ at 15 \AA, a wavelength
 accuracy of 5 m\AA, and bin size between 7 and 14 m\AA \ in the first
 order. 
\begin{figure}
\begin{center}
\resizebox{\hsize}{!}{\includegraphics{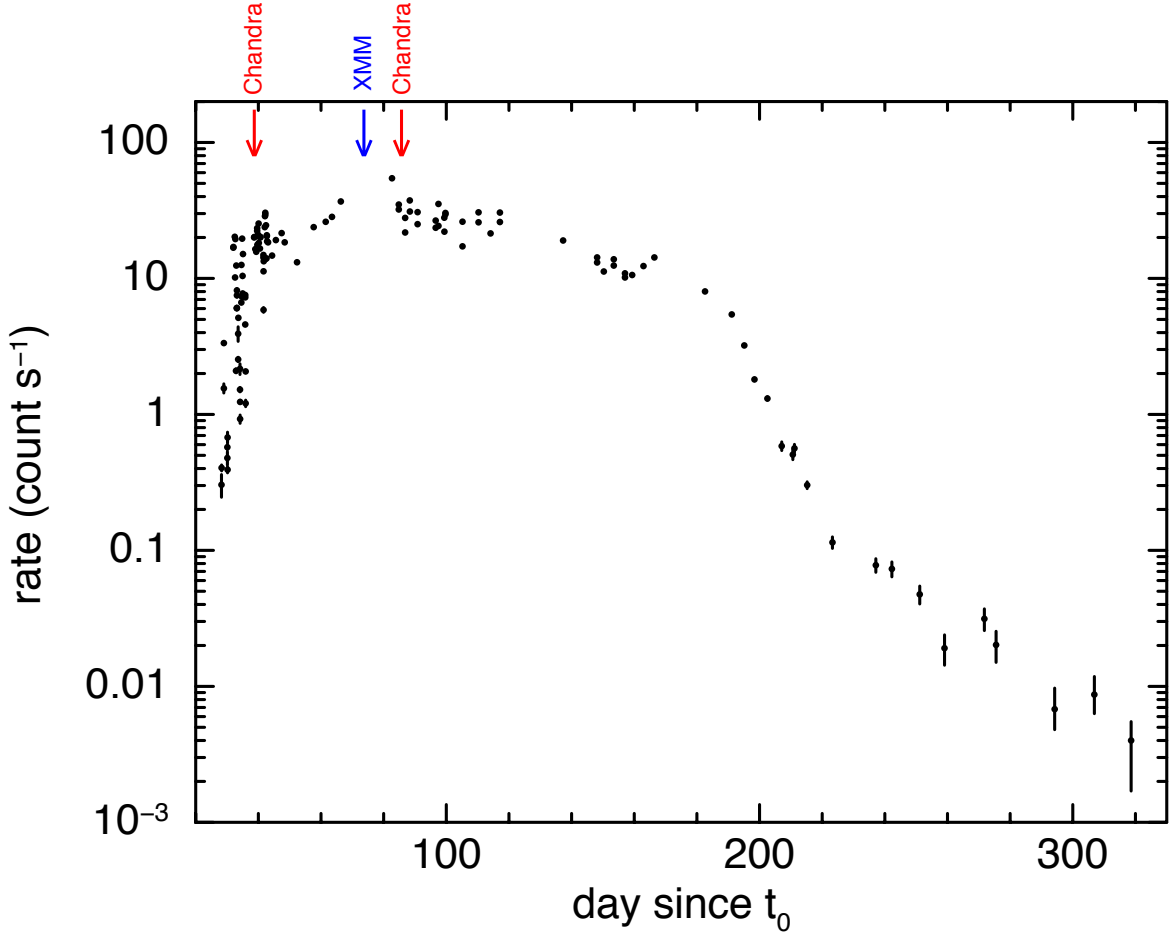}}
\end{center}
\vspace{-0.7cm}
\caption{ The {\sl Swift} XRT light curve in the 0.3-10 keV
 range published by \citet{aydi2018}; the dates of observations with
  {\sl Chandra} and {\sl XMM-Newton} are indicated. The initial time
  of optical maximum was 2016, October 9.}
\label{Kims}
\end{figure}
\begin{figure}
\begin{center}
\resizebox{\hsize}{!}{\includegraphics{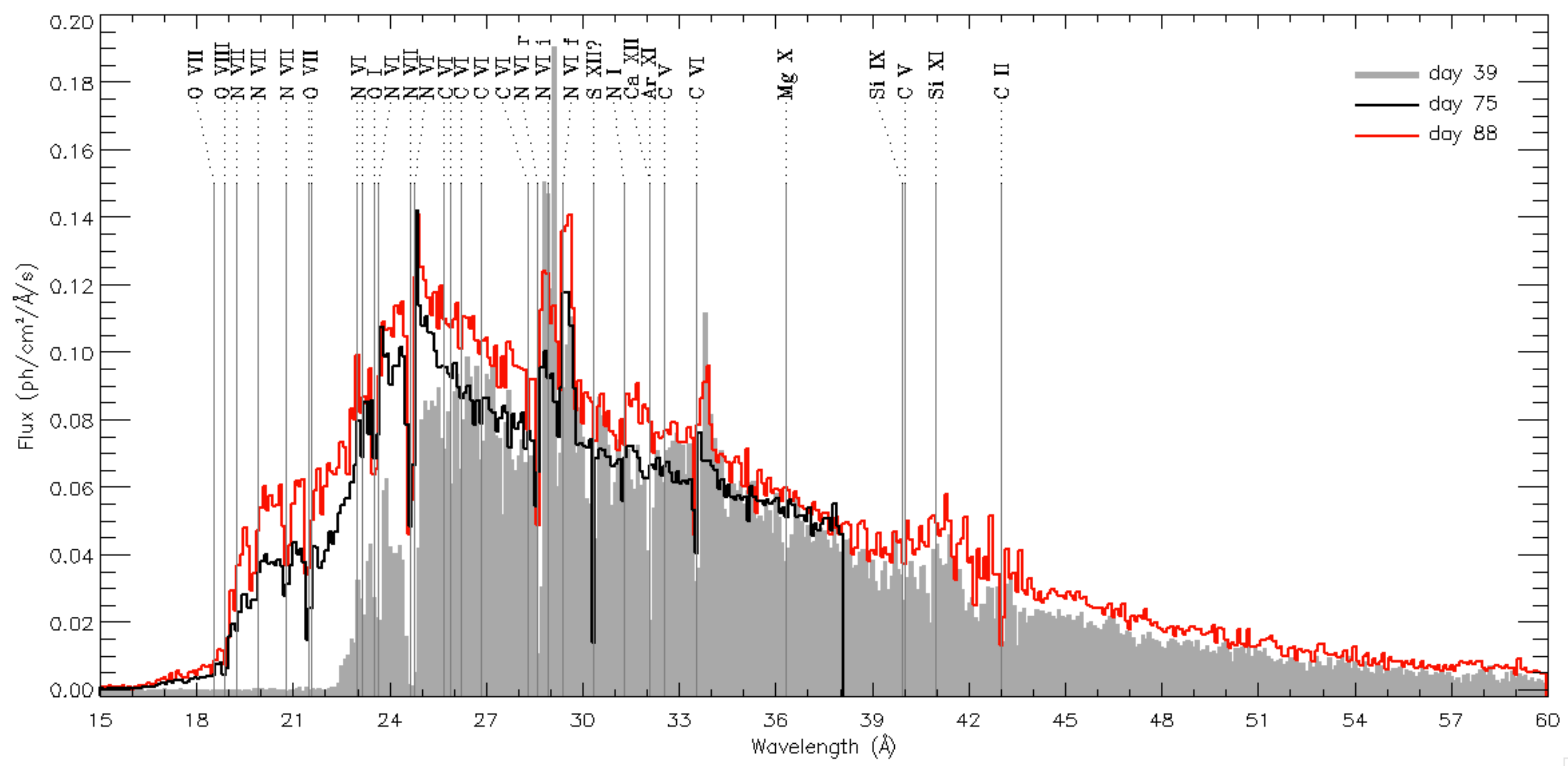}}
\end{center}
\caption{The flux-calibrated grating spectra of all three observations in units of counts of
 photons (ph) cm$^{-2}$ s$^{-1}$ \AA$^{-1}$.  We marked the wavelengths of the main 
 the absorption lines we identify, assuming they are not at rest, but blue-shifted by
 1800 km s$^{-1}$, except for lines of OI, NI and CII that are indicated at rest
 wavelength because we concluded that they are produced in the local interstellar medium.
  }
\label {Spectra1}
\end{figure}
\begin{figure}
\begin{center}
\resizebox{\hsize}{!}{\includegraphics{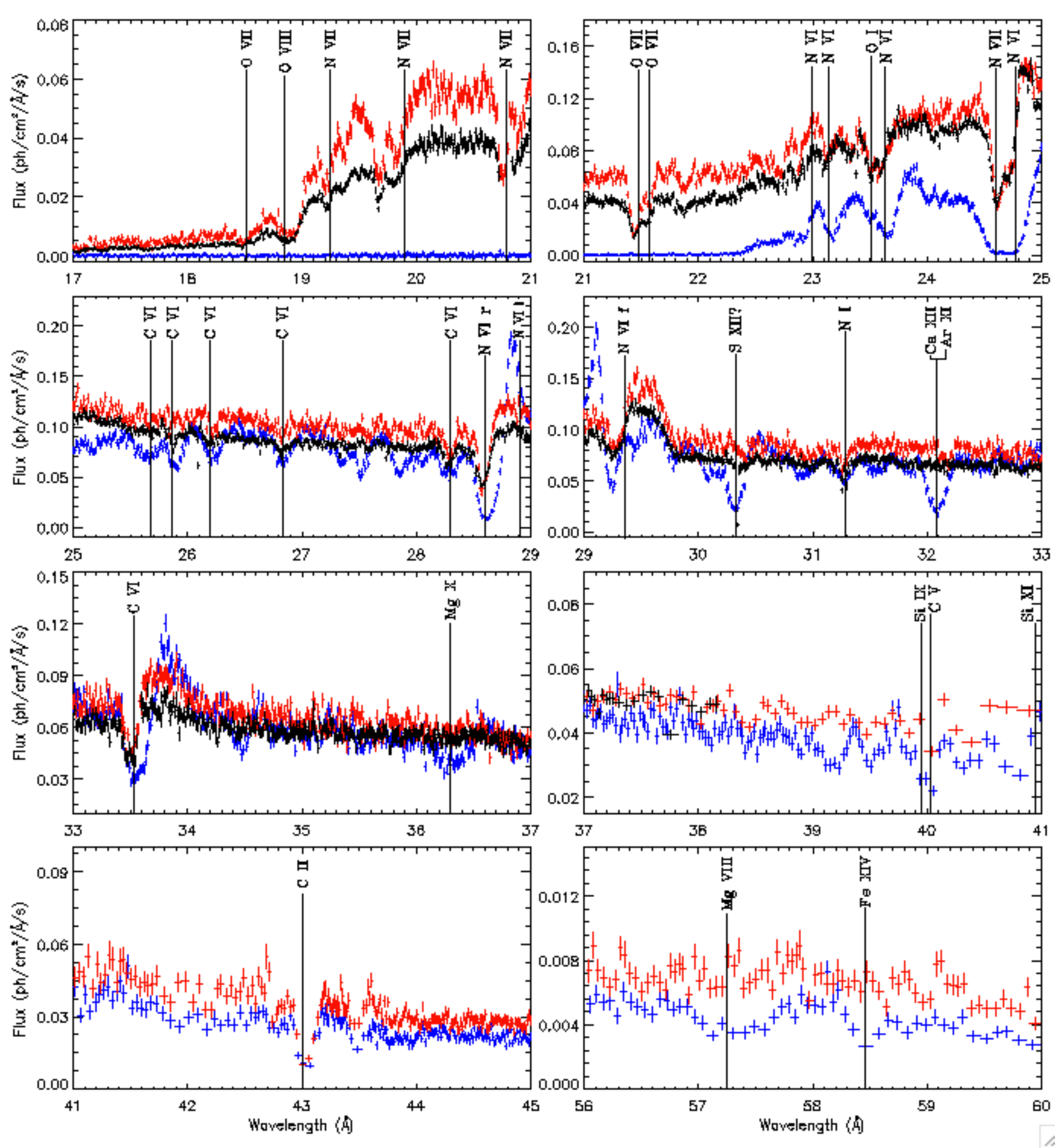}}
\end{center}
\caption{Details of the fluxed spectra in regions rich in absorption  lines,
 with proposed identifications, assuming the identified
 lines are blue-shifted by 1800 km  s$^{-1}$ (except for the interstellar medium
 lines of O I, N I and C II, which are marked at their
 rest positions).
 The spectrum of day 39 (November) is  plotted in blue, the one of day 75 (December) in black,
 and the one of day 88 (January) in red.  We plotted the error bars in order to show
 the data quality, and include here additional, more tentative
 identifications. In the last three panels we binned the Chandra data
 with S/N$\geq$10 or at least 10 counts per bin, 
 while the resolution is the instrumental one in the other panels
 for the Chandra data, and the XMM data were binned with at
 least  S/N$\geq$7 or 7 counts per bin.}
\label{Spectra2}
\end{figure}
 During the {\sl XMM-Newton} exposures all
 instruments were operated, including the Reflection Grating Spectrometers (RGS), 
 in addition to the EPIC pn and MOS cameras and the Optical Monitor (OM). 
 The EPIC-pn camera calibrated energy range is 0.15-10 keV and the EPIC-MOS range 
 is 0.3-10 keV; because of the large pile-up in the EPIC
 data, for this paper we made use only of the RGS X-ray data. 
  The OM data were obtained with the UVW2 filter,
 which has an effective wavelength of 2120 \AA \ and a width 
 of 500 \AA. The nova was observed in fast mode
 (yielding a count rate measurement every 0.5 s), in addition
 to the imaging mode that allows to calculate the magnitude
 integrating over much longer times. We could not directly compare 
 the magnitudes that were measured with the UVW2
 filter with the {\sl Swift} UVOT light curve
 presented in \citet{aydi2018}, because the UVOT exposures 
 were done only in a different bandpass filter
 at the time of the OM observation, but we find that the OM magnitudes
 are consistent with the trend of
 the UVOT light curve.  We used the UV OM lightcurve in
 fast mode for the timing analysis. 

\section{The high resolution X-ray spectra}
We extracted the {\sl Chandra} HRC-S+LETG spectra with their first order
 grating redistribution matrix files and ancillary response files with
 the CIAO 4.9 task {\it chandra{\textunderscore}repro},
 with the 4.7 version of the calibration
 package {\sl CALDB}. We coadded the positive and negative order spectra
 with   {\it combine{\textunderscore}grating{\textunderscore}spectra}
 to increase the signal-to-noise ratio. We did not find it necessary
 to correct for the higher spectral orders, which gave negligible
 contamination.

 We used the XMM-SAS software version 16.1.0 and ran the pipeline with
 the {\it rgsproc} task to obtain the spectra up to third order, but
 found only the +1 and -1 order spectra to be of interest. We
 combined them with  {\sl rgscombine} to obtain a higher signal-to-noise
 spectrum.

 The X-ray grating spectra are shown in Fig. 2
 and in more detail in Fig. 3, in which we have marked the
 strongest absorption lines. 
  We find absorption lines due to transitions of nitrogen and carbon  
 in all three spectra,  absorption lines of silicon, 
 magnesium, calcium and possibly argon and sulfur on day 39, and
  additionally, we detect H-like and He-like oxygen features
 in the spectra of days 75 and 88,  (indicating higher
 T$_{\rm eff}$).
 All the absorption lines that are consistent
 with the WD atmospheric origin are blue-shifted by about 1800 km s$^{-1}$ in the first
 observation, while the blue shift seems to vary a little
 for different lines, and reaches
 up to 2100 km s$^{-1}$ in the second and third observation. 
 We added a question mark to a line of S XII (rest wavelength 
 30.514 \AA) because  there is also a possibility that it is instead Ca XI (rest wavelength
 30.503 \AA, but a stronger Ca XII at rest wavelength 30.448 \AA \ should fall
 at 30.266 \AA \ and is not observed).
 A line that overlaps and may even be blended with other features is
 the one attributed to  Mg X (rest wavelength 36.518 \AA):
 it overlaps with a transition due to Ar XI, and is also very near a 
 Cl XI line (rest wavelength 36.518 \AA). 
 In order to finalize these identifications, a custom-tailored atmospheric model
 with ad-hoc, fine-tuned abundances will be crucial. 

  In addition to the blue-shifted atmospheric lines, like in  many other X-ray  spectra
 of novae we detect several 
 absorption lines, at rest wavelength, of transitions that occur with much
 lower ionization or excitation
 potential, namely O I (23.508 \AA), N I (31.28 \AA) and
 several features due to C II and C III around 42-43 \AA.
 To the carbon features we
 dedicate a separate section below, because they have only been identified
 and measured very recently in the spectra of Galactic novae 
\citep{gatuzzness}.  These spectral features and K-edges
 at low ionization are always observed at their rest wavelength
 and are thought to be produced in the interstellar medium (ISM)
 between the nova and us. 
 They are not typical
 of novae; they are detected whenever there is a bright X-ray source,
 including many low mass X-ray binaries and more rarely AGN, acting
 as  lamps giving a strong backlight 
 \citep[see][and references therein]{gat15}.

   A few features remain unidentified. These include several
 lines in the day 39 spectrum, which has the strongest
 absorption lines in the soft range: one at 
 25.93 \AA \ (accounting for blue-shift, it corresponds to
the unidentified 26.06 \AA \ feature of the list of
 \citet{ness2011}), two lines measured at 27.33 \AA \ and 27.5 \AA \ that are 
close, but not perfectly coincident accounting for the blue shift, with 
 unidentified lines of RS Oph \citep{ness2011},
and finally two lines at 30.87 \AA \ and $\simeq$38.1 \AA \
and $\simeq$39.2 \AA. 

\subsection{Comparison with other novae}
 While the oxygen, nitrogen and carbon absorption lines have been observed in  
 most other novae, in Fig. 4 we show that the SSS spectrum 
of day 39 appears extremely similar
 to the one observed at a later post-outburst epoch for KT Eri 
 (day 84 post-outburst), as taken from the Chandra archive.
 Another nova X-ray spectrum that presents
 similarities is that of V4743 Sgr at day 180  \citep{rauch2010},
 but there was less flux at the longest wavelengths 
 and there
 were other differences in the observed features. The lines 
 that we attributed to transitions of magnesium, argon and, possibly,  sulfur,
 are either missing or much less prominent in V4743 Sgr. \citet{rauch2010} 
 found solar sulfur,
 and depleted magnesium with respect to solar values in V4743 Sgr.
 Sulfur, argon, calcium and aluminum, are intermediate atomic weight
elements expected
 to be overabundant with respect to the solar value in oxygen-neon novae,
 but they should have solar abundance (or less) in novae on carbon-oxygen WDs
 \citep[see, e.g.][]{Starrfield2009}.
An accurate determination of the
sulfur abundance
 will be particularly important because the ratio of its abundance over that of oxygen
 and aluminum is one of   a few ``nuclear thermometers'' that constrain the degree of mixing
 of the accreted material with inner core elements \citep{Kelly2013}.

 The intermediate mass elements 
 are synthesized on oxygen-neon WDs, where neon-sodium and
 magnesium-aluminum cycles, that do not occur on carbon-oxygen white dwarfs, operate in
addition to the CNO cycle  \citep{Jose1998, Starrfield2009, Kelly2013}.
 Magnesium is abundant in a superficial layer on oxygen-neon WDs, and as a consequence 
 atomic magnesium (sum of the three
 isotopes $^{24}$Mg, $^{25}$Mg and $^{26}$Mg)  is expected above solar abundance in
 oxygen-neon novae \citep{Jose1998, Kelly2013}.  Thus we need to be able to model the
 spectrum sufficiently well
 to derive whether it is fitted with the range of abundances that would produce these
 lines, in order to be able to tell whether the underlying WD is of oxygen-neon or of
 carbon-oxygen. This ``first order''
 initial comparison indicates that V4743 Sgr, KT Eri and N SMC 2016
 at this stage in the evolution
 of each nova had approximately the same  T$_{\rm eff}$: it was
 estimated to be 740,000$\pm$70,000 K
 in this spectrum of V4743 Sgr and our preliminary
 result for KT Eri also indicates T$_{\rm eff}$ around
 700,000 K.
 However,  T$_{\rm eff}$, which at its maximum is an indication of the WD mass 
\citep{starrfield2012, Wolf2013}), was still raising to the peak in N SMC 2016.
  Both  novae whose
spectra  we  plotted in  Fig. 4, N SMC 2016
 analysed in this work and KT Eri, are likely to host oxygen-neon novae. V4743 Sgr,
 instead, fitted  by \citet[][]{rauch2010}
 with a model for a carbon-oxygen WD, with solar abundances 
 of intermediate atomic mass elements and depleted magnesium,
 very likely occurred on a carbon-oxygen WD.
\begin{figure}
\begin{center}
\resizebox{0.87\hsize}{!}{\includegraphics{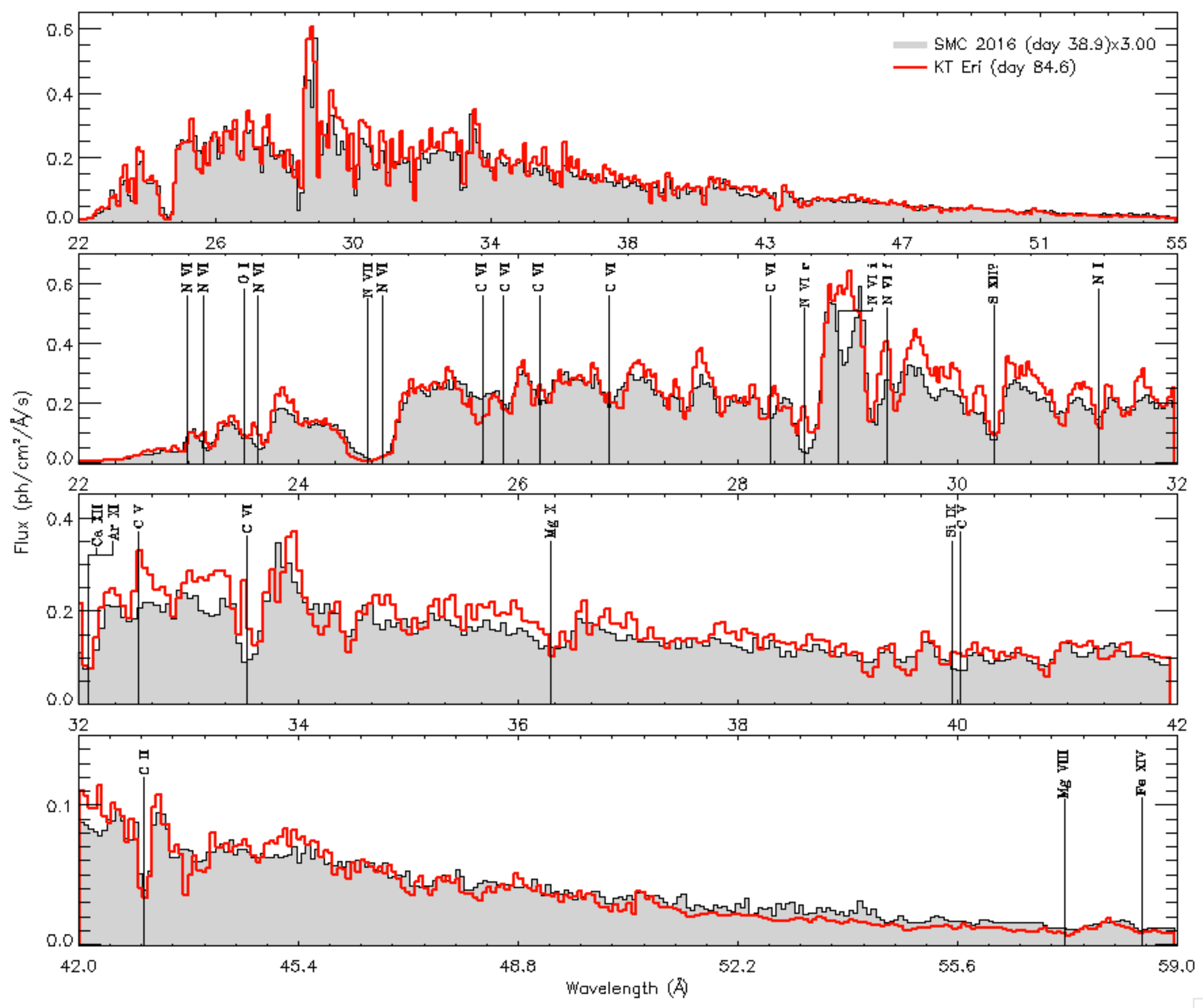}}
\end{center}
\caption{Comparison of the Chandra HRC-S+LETG spectra of N SMC 2016 at day 39
 and KT Eri at day 84. Several absorption lines (assumed
 to be blue-shifted by 1800 km s$^{-1}$)
 are also indicated, together with the rest wavelength positions
 of the ISM lines identified above.}
\end{figure}

\subsection{Model fitting}
 In order to classify the nova correctly and estimate the abundances in the hot atmosphere,
 including verifying the possible oxygen-neon WD nature,
 a physical model is necessary. 
 For this article we experimented with two different models: the Non Local Thermodynamical
 Equilibrium (NLTE) model atmosphere TMAP \citep{rauch2010} and the synthetic models
 for expanding atmospheres ``wind-type'' WT model of \citet{vanRossum2012}.
 Since it takes a considerable effort to produce new model grids with different
sets of abundances, we used the already available grids in the above papers, in order
 to understand whether these models are at all viable for
 this nova. A detailed abundance study requires
 extending the grids of models to find the parameters that reproduce 
 our spectra; additional work is in progress and will be the subject
 of a forthcoming paper (Rauch et al. 2018, in preparation).
 
A basic question we
 would like to answer is: given the blue shift of the absorption features, what is the correct way
 to model this WD atmosphere from which an outflow is still present?
 The radiation driven wind models \citep[see,e.g.][]{hachisu2007}
predict that the end of mass
 loss occurs around the time the WD appears as a SSS, but in the SSS we still
 find blue shifted absorption lines in this and in other novae.
  A typical case is RS Oph, as it 
was analysed by \citet{hachisu2007}  and the end of mass loss was inferred,
 while 
 the X-ray spectra measured by \citet{ness2007, nelson2008}
 still showed blue-shifted absorption features.
  Although the pseudo-photosphere of the nova
 has clearly {\it contracted}, as foreseen by all the models,
  and the WD radius has shrunk
 from that of a red giant configuration
 at the time of the optical luminosity peak, 
 becoming very compact,  as it is predicted by the models 
 for the constant bolometric luminosity phase
\citep[e.g.][]{starrfield2012, Wolf2013},
the blue-shifted absorption lines
 are explained if there is still 
an outflow from an expanding atmosphere.  This residual, late-phase
 outflow has not been predicted by any models, but 
 the blue shift of the absorption features
 has  indeed been detected in most SSS high resolution spectra 
 of novae \citep[see
 among others][]{rauch2010, ness2011, orio2012, orio2013}. It
 is interesting to note that the lines are often 
 blue-shifted with about the same velocity observed in the emission lines
 of the optical spectra of the outburst.
 
 When the WD photosphere has shrunk
 back to an almost ``normal'' WD radius, the amount of mass
 lost in the wind is likely to be very small. With
 a large mass outflow, the absorption features would not be so deep and
 the profile  would not remain similar to the static atmosphere case,
therefore we reasoned
 that a first attempt at a fit can be done  
 with a static atmosphere. Most novae
X-ray grating spectra are in fact very similar
 to CAL 83, a steadily burning source that does not present mass loss,
 albeit generally the novae we have observed are at
 higher T$_{\rm eff}$ than CAL 83,
  which was found to have T$_{\rm eff} \simeq$ 550,000 K \citep{Lanz2005}.
The  TMAP code nova model grid was initially developed
 with the aim of reproducing the abundances of V4743 Sgr
 \citep{rauch2010}; the published grid includes various combinations
 of highly non-solar abundances as calculated for 
 the burning layer in models of a carbon-oxygen WD's by Prialnik and
 coauthors, obtained directly from them). The other grid of models
 we used was obtained with the ``wind type'' WT expanding atmosphere
 code, which includes a larger number of atomic species, but has only solar abundances
\citep{vanRossum2012}.

In order to take the intervening
 column density into account, we used the T\"ubingen-Boulder absorption model
and performed the calculation with the relative routines 
TBNEW in IDL and TBABS in XSPEC \citep[see][]{Wilms2000}.
 In Fig. 5 we show the TMAP
 fits for all observations, whose parameters are given in detail
 in Table 2, and in Fig. 6 we present the WT fits.
 The {\sl Chandra} spectra,
 providing a softer range, allow us to evaluate the absorbing column
 density of equivalent neutral hydrogen N(H) better
 than the {\sl XMM-Newton} one. We note that \citet{aydi2018}, referring
also to \citet{haschke2012}, adopted a value A$_{\rm V}$=0.11$\pm$0.06, 
 corresponding to a very low 
column density, N(H)=1.7 $\times 10^{20}$ cm$^{-2}$ \citep[following][]{Predehl1995}. 
 Because in X-rays we are observing deeply inside the shell, there
 may be additional column density, but it has to be low enough
for the material to be optically thin to supersoft X-rays.
Despite the unexplained velocity of the absorption lines,
 we obtained a much better fit to the observed spectra with TMAP 
 (with the caveat of artificially shifting the absorption
 features to match the data, see Fig. 5),  than with the WT models 
 (see Fig.6).

\begin{table*}
\caption{Parameters of the best fit obtained for N SMC 2016 with
 TMAP. The unabsorbed flux is calculated in the 12-124 \AA range \ ($\approx$0.1-1 keV)
 and represents more than 95\% of the bolometric flux.}
\begin{center}
\begin{tabular}{rrrrrr}\hline\hline \noalign{\smallskip}
Day & T$_{\rm eff}$ (K) & N(H) (10$^{20}$ cm$^{-2}$) &  F(unabs) (erg/s) & [N/N$_\odot$] & [C/C$_\odot$] \\ 
\hline
 39 & 716,000 & 4.92 & 2.16  $\times 10^{-9}$ & 1.803 & -1.513 \\
 75 & 852,000 & 5.44  & 2.72 $\times 10^{-9}$ & 0.937 & -0.529 \\
 88 & 904,000 & 4.04 & 2.59 $\times 10^{-9}$ & 1.159 & -0.596 \\
\hline  \noalign{\smallskip}
\end{tabular}
\end{center}
\end{table*}

 The TMAP fit was done in XSPEC.  The best fits we obtained,
 tested according
 to two methods of analysis, the $\chi^2$ and the Cash statistics \citep{Cash1979},
  are shown in Table 2. These fits yield a temperature 716,000 K for day 39, 
 852,000 K for day 75 one and 904,000 K
 for the last observation of day 88 done with {\sl Chandra}.  The fits are not
 perfect, but they approximate the shape of the continuum and the 
 strongest absorption features quite well. 
 However, the fits yield  
 excess flux both on the soft portion of the spectrum for the spectra 
 of days 75 and 88, and also excess flux on the 
 ``hard'' side for day 39. For  day 39, 
 if  we decrease the temperature
 from 716,000 K to 650,000 K, the N VII K-edge
 that is responsible for cutting the hard
 X-ray flux shortwards of 18.59 \AA \ is matched,
 but at the expense of under-predicting 
 even more the soft flux longwards of
 45 \AA. Decreasing the value of the absorbing column N(H) 
 does not lead to a perfect fit in this soft part of the spectrum.
The main problem here is
 that the grid step of 50,000 K results in too large an uncertainty in T$_{\rm eff}$
 to have sufficient precision to match the absorption
 edges perfectly.  The {\sl Swift} XRT
 trend suggests that during the day 75 observation the temperature may
 have still been lower than on day 88,
 but because of the large step in  T$_{\rm eff}$  in the grid,
 we can only constrain it to be in the 850,000-900,000 K range for both days.
 We note that the model misses a feature of He-like nitrogen
  at about 28.78 \AA \  that is 
still observed at days 75 and 88. This
may mean that we have overestimated the temperature for
 lack of a suitable grid step, or that the best
 fit underestimates the nitrogen abundances.
 For a peak SSS
 temperature in the ballpark  of
 the value we obtained, the WD mass is 1.2$\leq$m(WD)$\leq$1.3 M$_\odot$
 following \citet{Wolf2013} and 1.25 M$_\odot$ following \citet{starrfield2012},
 consistently with the values inferred by \citet{aydi2018}.  

  In the publicly available grid of models, the parameters
 that mainly vary are the carbon and nitrogen abundance. While
 on day 39 the best fit was obtained with the highest ratio of  N/C
 (in mass
 relative to the solar value), the ratio was lower on days 75 and 88,
and
this may imply that the atmospheric layer was mixing with newly accreted material
 from the companion.    
At a distance of 60 kpc, the unabsorbed flux  on
 day 38 implies an X-ray luminosity of 9.26 $\times 10^{38}$
 erg s$^{-1}$ in
 the 12-124 \AA \ range (equal to almost all the bolometric luminosity)
 and of 1.11 $\times 10^{39}$ erg s$^{-1}$  on days 75 and 88.
 The fits indicate only a modest increase in unabsorbed flux from day 39 to day 75,
 consistently with the WD  photospheric radius still shrinking at approximately
 constant bolometric luminosity.
 We notice that these
 values are above Eddington level and indicate
 a larger radius than that of a WD of 1.25 M$_\odot$.  After
 the 40 days of super-Eddington luminosity inferred in the optical and UV range,
 the nova luminosity exceeded the Eddington limit 
 at least until day 88 after having shifted the peak
 of emission to the X-rays; because the {\sl Swift} X-ray Telescope light curve
 \citep[Fig.1, and][]{aydi2018} implies that the SSS did not decline,
 and remained close to the level of day 88, the luminosity must have
 been above the Eddington level for even over 5 months
 after the outburst. This implies that the stellar
 configuration was not stationary yet, and is consistent with the outflow indicated
 by the blue-shifted lines.

 \begin{table*}
\caption{Parameters of the most acceptable we obtained for N SMC 2016 with
 the WT model.  For days 75 and 88,
 the temperature is the highest available in the grid, because no model was calculated
 with higher T$_{\rm eff}$. We assumed a 60 kpc distance to the nova.}
\begin{center}
\begin{tabular}{rrrrrr}\hline\hline \noalign{\smallskip}
Day & T$_{\rm eff}$ (K) & N(H) (10$^{20}$ cm$^{-2}$) &  v$_{\rm ej}$ (km s$^{-1}$  
& log(g) & $\dot m$ (M$_\odot$ year$^{-1}$) \\
\\
\hline 
 39 & 650,000 & 5.00 & 1800 & 8.9 & 7.6 $\times 10^{-9}$ \\
 75 & 700,000 & 6.00  & 1800 & 9.01  & 7.6 $\times 10^{-9}$ \\
 88 & 700,000 & 6.00 & 1800  & 9.01 &  7.6 $\times 10^{-9}$ \\
\hline \noalign{\smallskip}
\end{tabular}
\end{center}
\end{table*}

 It is interesting to notice that the absorption
 features of the nova, though blue-shifted, are as deep as predicted by the static atmosphere
 model. In the day 39 
 spectrum, the  N VII 24.78 \AA \ line appears even saturated. A likely
 explanation is that little mass is flowing out, because
 only in such a case does the WT model become more
 similar to a static one, and this would be the reason the static model
 approximates the observed spectrum reasonably well.

  Because the WT model is not available
 in the standard HEASOFT XSPEC software or in other spectral
 packages, we calculated the convolution of the flux at
 each wavelength with the absorption (or transmission) function, by using
 IDL. In a static model the effective
 gravity is dependent on the temperature T$_{\rm eff}$,
 but in the WT model, there are more parameters (the velocity
 and the mass outflow rate $\dot m$),   but
 the parameters are not independent  on each other.
 For a given temperature and effective gravity, the
 mass outflow rate is determined by the velocity of the wind, but 
 in the WT model the wind velocity does not correspond
 to the observed blue shift of the absorption features,
 because the lines are produced in an extended, outflowing medium with a complex structure.
 Since the model is not included in a spectral fitting package, we first
fitted the data 
 with different sets of parameters in the grid, then compared them by eye. 
  The fits we found most acceptable are plotted in Fig. 6.
 There is a sufficient difference
 between certain sets of models that we could rule out 
 a large number of  them, namely those with high mass outflow rate
 and low effective gravity. Thus, we focused on
 fits with ``dlg-03'' and ``dMv442'' series, which
 includes the  combination of the 
 highest effective gravity and the lower mass outflow rate for a given velocity.
 Because the models predict the absolute flux, and we know the distance
 to the source,
 we found that the temperature that is necessary to match the observed
 flux produces too hard a spectrum. The particular structure of the medium
 in which the transitions occur in the WT models 
 smears the absorption edges and cannot reproduce them.
 An ``experiment'' with an XSPEC blackbody model with overlapping
 absorption edges showed us that
 the latter are very sharp even with solar abundances, therefore
 we suggest 
 that it is the wind structure that cancels, or rather
 ``velocity-smears'', the absorption edges
 in the WT model, not the lack of enhanced
 abundances. \citet{vanRossum2012} found that he could fit one
 of the V4743 Sgr spectra with systematically
 lower temperature than the static atmospheric models, but in this case 
 we cannot lower  T$_{\rm eff}$, or else the flux would be 
 so low that the nova should be in the Galaxy (this has been was ruled out
 by  \citet{aydi2018}).  
 Thus, in practice, the high temperature is constrained 
by the absolute luminosity.  For instance, for the first
 spectrum a fit with T$_{\rm eff}$=450,000 K would imply a total absorbed luminosity
 lower than observed by two orders of magnitude at a distance of 60 kpc. The nova
 would have to be Galactic, but even with this assumption, we still do not
 obtain any match with the observed spectral features. We note that, 
 even if we match the flux
 to fit a distance of only 45 kpc that was not completely ruled out 
 by \citet[][]{aydi2018}, the  T$_{\rm eff}$ 
 indicated by the model for the first observation would not be lower than 650,000 K. 
 The highest temperature in the grid is 750,000 K, and  in the middle
 and lower panels of Fig.  6 we show how it fits
 the second and third observations (poorly, however, clearly it is only
 a lower limit).  The parameters of the WT
fits are reported in Table 3.
 
The WT models also fail to reproduce the observed blue-shift and depth of the lines. 
   In fact, as the surface gravity decreases and the density
 structure changes in the wind with respect
 to a static atmosphere, the absorption features tend to become much shallower  and are
 at times accompanied by emission wings in P-Cyg profiles, causing only a modest blue-shift of
 the absorption features that does not equal the wind velocity. 
\begin{figure}
\begin{center}
\resizebox{0.95\hsize}{!}{\includegraphics{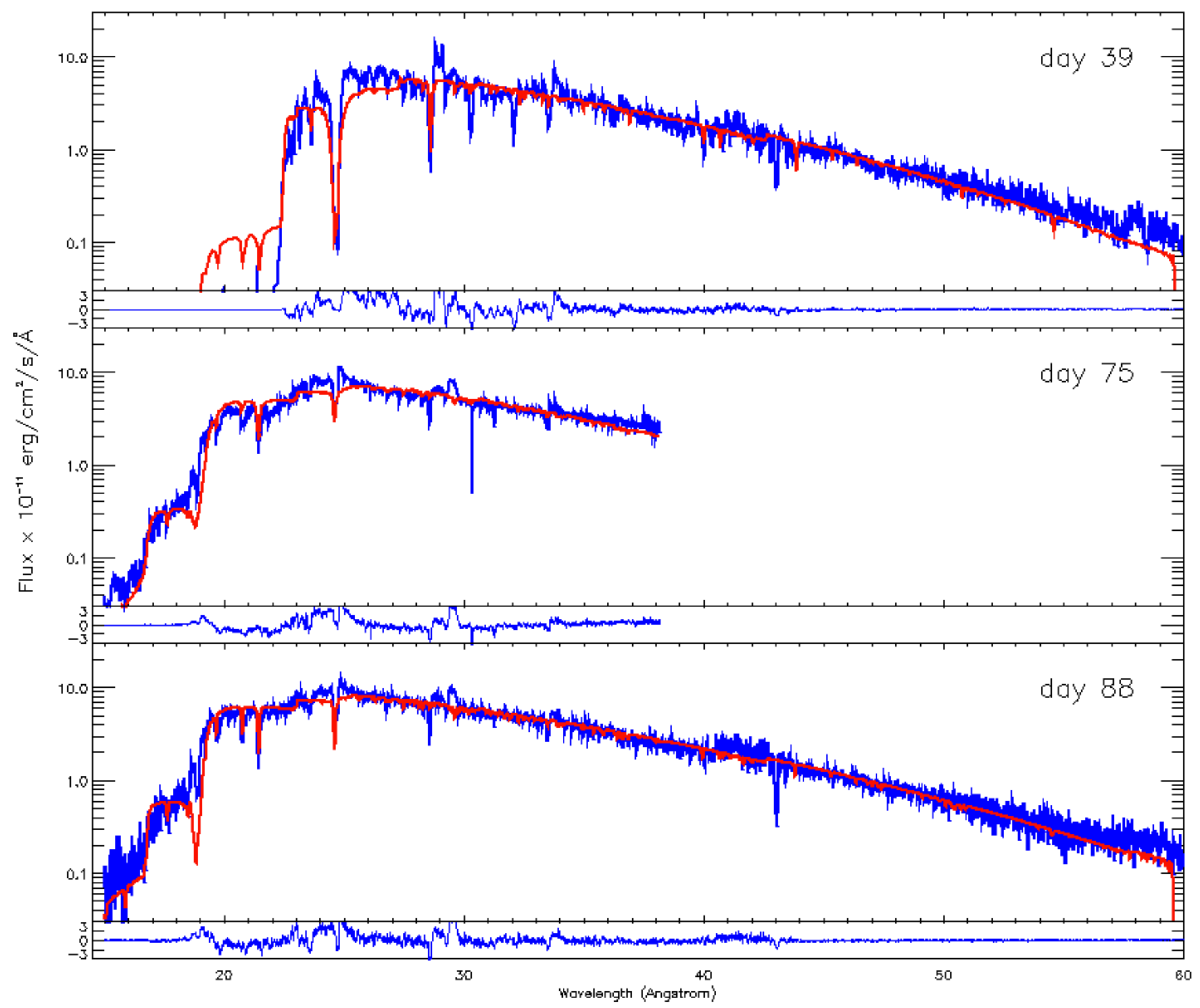}}
\end{center}
\caption{In the top panels,
the observed spectra of N SMC 2016 
taken on day 39 and 88 with the Chandra HRC-S+LETG setup, and on day 75
 with the RGS of XMM-Newton, are traced in blue  for each date. The 
 red lines show the XSPEC best fit with a TMAP model with log(g)=9 and
the parameters in Table 2.  The fit has been obtained by artificially moving the
 atmospheric absorption lines with respect to the original model, in order to match the observed
 blue shift.  For each spectrum, the lower panel shows, in linear scale,
 the residuals, namely the difference between the data and the model.}
\end{figure}
\begin{figure}
\begin{center}
\resizebox{0.95\hsize}{!}{\includegraphics{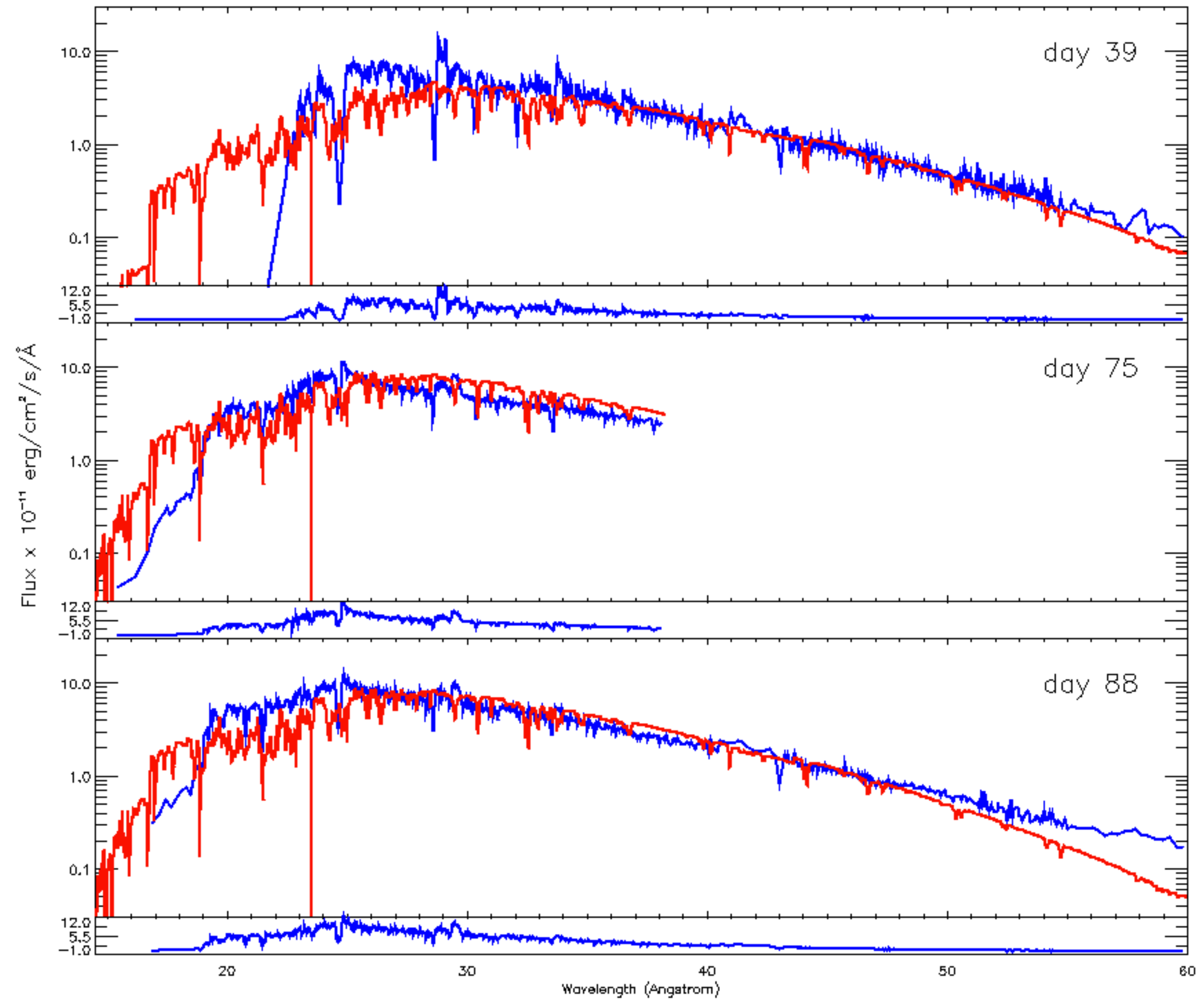}}
\end{center}
\caption{
The observed spectra, as in Fig. 5, are traced in blue, 
 and in the top panels
 for each observation, in red we trace the most acceptable fit we were able to obtain 
 with the  WT model, using  the parameters given in Table 3. We assumed
 a distance of 60 kpc to normalize the flux.  For each spectrum, the lower panel shows,
 in linear scale,
 the residuals, namely the difference between the data and the model.}
\end{figure}

There are structures in the observed  spectra
 that appear to consist of emission lines,
 at least  around 29 \AA, between 33 and 34 \AA \ and possibly also
 a little
 redward of about 43 \AA, that no models can reproduce (we
 note, however, that the LETG spectrum is noisy around 43 \AA, 
 because of the very low effective
area, due to instrumental (HRC-S) absorption by the carbon
K edge at 43.6 \AA). 
  A combination of static atmosphere and plasma in collisional ionization equilibrium 
 in the ejecta has been found to be present in several
 novae: U Sco \citep{orio2013}, T Pyx \citep{Tofflemi2013}, V959 Mon \citep{peretz2016},
 but some of the apparent emission lines in N SMC 2016 are in the
 range below 40 \AA, and are not reproduced with the plasma temperature 
 of the available ``packaged''
 data analysis models of collisional ionization equilibrium. 

  Our  proposed future course of action to reveal the chemical composition
 of this nova WD is to first obtain a static model that fits the continuum and the
 absorption features better than the existing ones: this is currently
 being tried with new sets of parameters in the TMAP model (Rauch et al. 2018,
 in preparation). Only after this is achieved, we will try to analyse
 whether photoionization or collisional ionization explain the residual emission
 lines, and whether they may originate in the ejecta, far from the WD,
 or instead are connected with the residual wind from the WD.

\subsection{Lines originating in the interstellar medium}

 In Fig. 2, 3 and 4, we marked the lines of O I and N I at rest wavelength, because these
 are known to be  typically produced in the interstellar medium
  and to appear in the X-ray grating spectra of many different
 X-ray sources, both Galactic
 \citep[usually low mass X-ray binaries, see][]{gatuzz2016, gatuzz2018}, and extragalactic
 \citep[AGN, see e.g.][]{Nicastro2016a, Nicastro2016b, gatuzz2018}.
  For these reasons, and also because these lines are produced
 with a photoionizing source of much lower temperature than 
 needed for all other lines in the spectrum,
 we ruled out they originate in the nova and are intrinsic to it.

\begin{figure}
\begin{center}
\includegraphics[width=8.5cm,height=12cm]{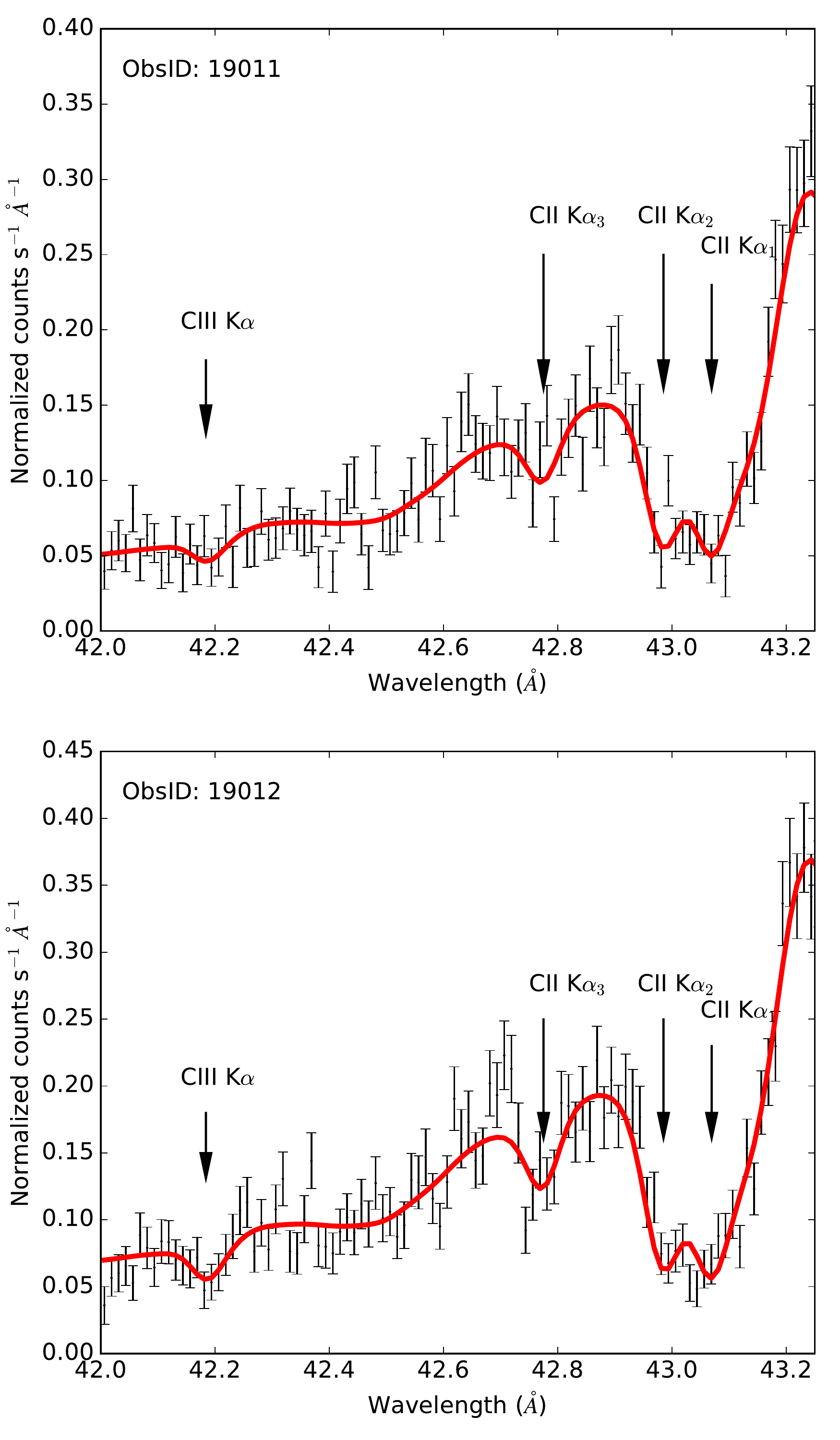}
\end{center}
      \caption{The red
 line shows the best fit results for the {\rm C}~{\sc ii} K-edge wavelength
 region, 
 including the atomic data benchmarking correction \citep{gatuzzness}.
 The top panel presents (in black) the day 39 data, the lower panel shows
 the data of day 88.
 }\label{fig_c2}
   \end{figure}

 Thanks also to recent updates in the calibration package {\sl CALDB} that better account
 for the HRC 43.6 \AA \ {\rm C}~{\sc i} absorption edge, making
 the calibration at the nearby wavelengths more secure, we were able to identify
   also
{\rm C}~{\sc ii}  and  {\rm C}~{\sc iii} features in the 38-44 \AA \ that have been recently    
  detected and measured in the ISM by \citet{gatuzzness},
 using Galactic novae as lamps. These authors (and several others quoted
 in their paper)  did not find evidence of  
{\rm C}~{\sc i}  features in the ISM, implying that neutral carbon has a very low column density, less that
 10$^{14}$ cm$^{-2}$ along multiple lines of sight.  However, \citet{gatuzzness}
were
 able  to detect several features of  {\rm C}~{\sc ii} 
 and {\rm C}~{\sc iii}, by using the {\sl Chandra} HRC+LETG spectra of four Galactic
 novae, and the {\sl ISMabs} model \citep{gat15}.
The spectrum of Nova SMC 2016 offers the possibility
 to calibrate the column density of these ions along
 a new line of sight, away from the Galactic center.  We identified the
 {\rm C}~{\sc ii} and {\rm C}~{\sc iii} K$\alpha$ triplets as well
 as {\rm C}~{\sc ii} and {\rm C}~{\sc iii} K$\beta$ resonances. The
 high-resolution achieved by the {\it Chandra} HRC-S+LETG instrument allows
 a detailed analysis of such features. 
We have not identified {\rm C}~{\sc i} absorption lines in the spectra
 of N SMC 2016. In Figure~\ref{fig_c2} we zoom into this specific spectral region
 for a clear understanding of these features, and  
 Table~\ref{tab_c2} lists the column densities obtained for each observation.
 Additional,
future observations of X-ray luminous novae located in the Magellanic Clouds,
 at different T$_{\rm eff}$, will be very useful
 for the estimation of the ISM chemical composition  in
 these lines of sight (e.g. ionization fractions, abundances and column densities).

\begin{table}
\caption{\label{tab_c2}ISM carbon column density best-fit results.
 The units are  $10^{16}$~cm$^{-2}$.}
\centering
\begin{tabular}{ccccccccccccccc}
\hline \hline
 Day & $N({\rm CII})$ &  $N({\rm CIII})$     \\
\hline
38   &$8.75  ^{+     1.28    }_{-    1.44    }       $&$     0.29^{+ 0.45}_{-0.26} $   \\
88   &$10.28 ^{      +1.35   }_{-    1.22    }       $&$     0.59^{+ 0.50}_{-0.34}   $   \\
\hline 
\end{tabular}
\end{table}

\section{Timing analysis}
\begin{figure}
\begin{center}
\includegraphics[width=11.5cm]{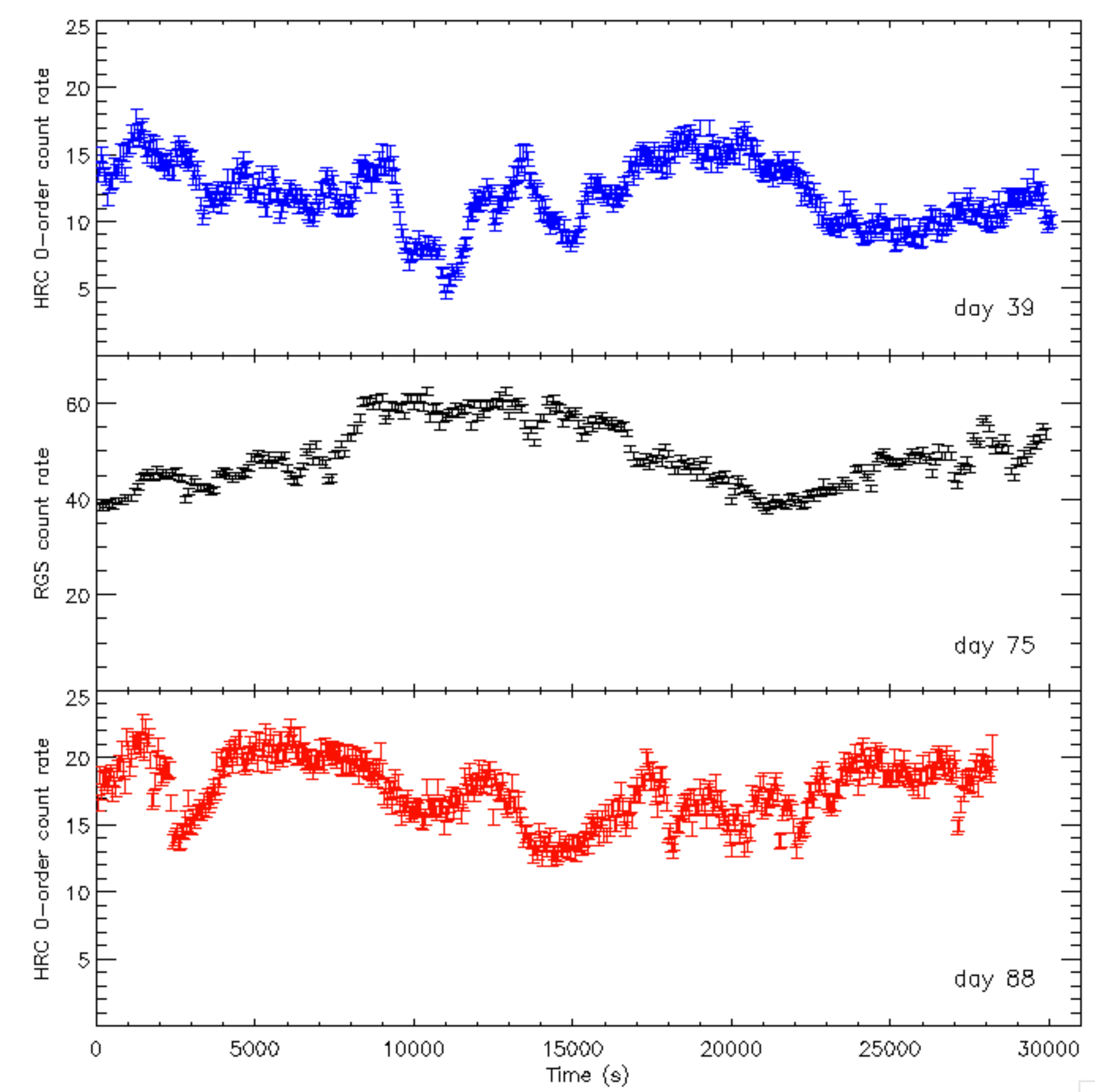}
\end{center}
\caption{Light curves measured with the {\sl Chandra} HRC-S camera
 as ``zero order'' in the 15-80 \AA \ on days 39 and 88, 
 and with the {\sl XMM-Newton} RGS (+1 and -1 orders, added)
 on day 75.  For plotting purposes, we have binned the HRC
 data every 50 s, the RGS data every 100 s (we used 
 different binning in the data analysis, see text).}
\label{lcs}
\end{figure}

 Fig.~\ref{lcs}
  shows the light curves measured in the three observations. The {\sl Chandra} light curves
 are the zero-order light curves measured with the
 HRC-S camera. The {\sl XMM-Newton} light curve is the RGS first
 order one, which, despite a 
 time resolution of only 5 s, is not affected by pile-up like the EPIC
 light curves. We restricted our analysis to the gratings' observations. 
There are  large irregular fluctuations: the count rate varies by factors of 
 3 on day 39, 1.5 on day 75 and almost 2 on day 88.   
We also extracted both {\sl Chandra} light curves of
 days 39 and 88 between 15 \AA \ and 33 \AA \ and compared them with the light curves
 between 33 \AA \ and 60 \AA. We also
 compared the light curves of day 39 split
 differently, in the 20-32 \AA \ range and in the
 32-50 \AA \ range,
  but we did not find that the modulations
 were larger, with stronger dimming, in the softer bands, as expected if 
 the fluctuations are caused by variable absorption. We repeated the experiment with
 the {\sl XMM} light curves of day 75 comparing the 15-27 \AA \ wavelength range 
 lightcurve
 with the 27-38 \AA \ lightcurve, but also in this case there was no indication of
 larger fluctuation in the lower energy band. We also extracted two separate
 spectra for day 75, one for count rate above 50 cts s $^{-1}$ and one for
 count rate below this level, and measured no significant difference in the
 spectra except for the continuum flux level.  

The variability appears to occur on different time scales,
 but we could not find an obvious coherent periodicity. 
 We performed a periodogram calculation with the
Lomb-Scargle algorithm \citep{scargle1982}, because it is suitable for data
 that are not equally spaced in time.
 The {\sl XMM-Newton} has high background or other ``flagged''intervals that could not
 be used, causing gaps in the light curve, so this algorithm
 is well suited. 

For a more detailed timing analysis, we used
power density spectra (PDS) instead of
 standard periodograms. This approach is suitable
to study quasi periodic oscillations (QPOs)
 or red noise. We estimated the PDS as in \citet{dobrotka2017}, i.e. we 
 split the light
 curve into $n_{\rm div}$ subsamples  (subdivision in time ranges),
  then we calculated periodograms in log-log space for every subsample,
 averaged them and finally binned the data in equally spaced bins if a minimum number of points per
 bin is fulfilled (otherwise the bin is larger).
Usually, the lower frequency end and the frequency
 resolution of the periodogram are determined by the length of the exposure, and the high
 frequency end by the Nyquist frequency. For our analysis,
instead of the latter we chose the frequency at which the periodogram trend appears
 to become almost constant, indicating pure white noise. The standard periodograms do not
 show any significant peaks, suggesting that there is no coherent periodicity.

\begin{figure}
\begin{center}
\resizebox{0.8\hsize}{!}{\includegraphics[angle=-90]{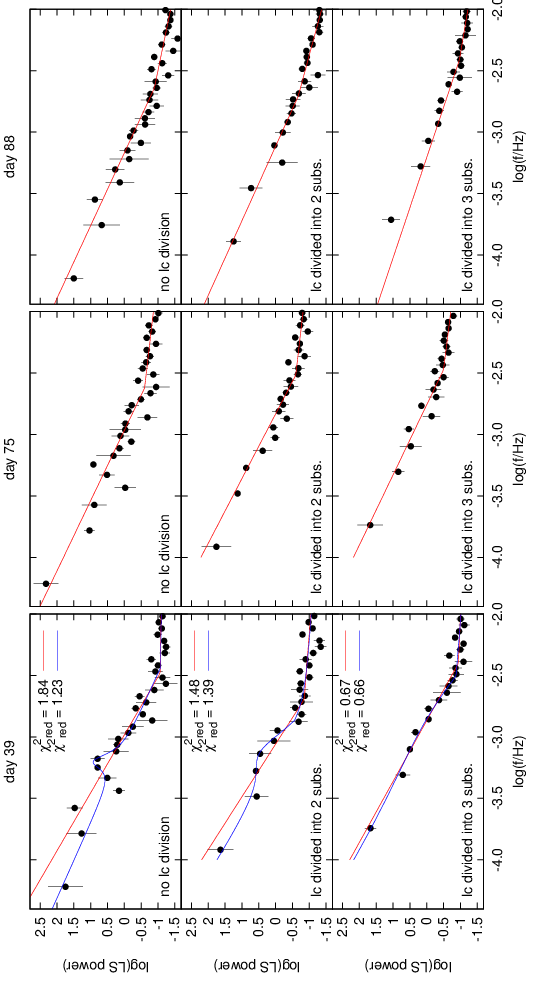}}
\end{center}
\caption{For all three light curves, we show here the PDS estimates 
 obtained with different intervals in which the 
 the light curves were split. The data points are averaged values per 
 frequency bin, the
error of the mean is our uncertainty estimate.
 The red lines show a broken power law fit,
 and evidentiate  the red noise below
log($f$/Hz) = -2.5. The blue line in the panels
 on the left shows the same broken power law
fit with an additional Lorentzian,
 used to describe a possible power excess below
log($f$/Hz) = -3.0 on day 39, 
 illustrating the effect of the light curve splitting
procedure.}
\label{pds}
\end{figure}

 Fig.~\ref{pds} shows the PDS calculated using
all three light curves, with $n_{\rm div} = 1$, 2 and 3.
 Note that the light curve of {\sl XMM-Newton} is binned with every 10 s, the
 first {\sl Chandra} light curve every 16s, and the second
 {\sl Chandra} light curve has 50 s bins.
 The trend is linear, suggesting
 that in all three cases we are only
 measuring red noise. This trend is shown by fitted broken
 power law fits\footnote{Two linear functions with different slopes before and after break frequency, in log-log space.}
 (red lines).

\begin{figure}
\begin{center}
\resizebox{0.7\hsize}{!}{\includegraphics[angle=-90]{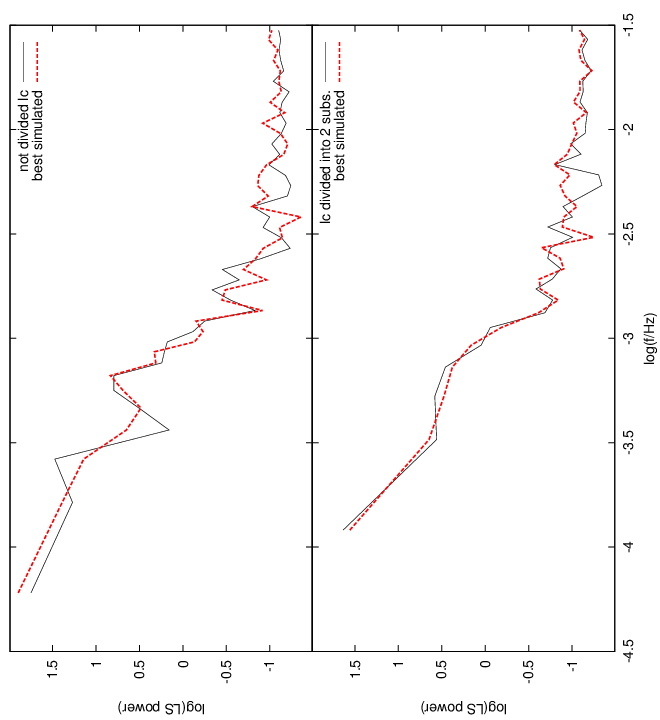}}
\end{center}
\caption{Observed vs simulated PDSs (best cases  chosen
 among 10000 simulations),  referring to the upper and middle
panels in the middle column of Fig.~\ref{pds}, only for day 39.}
\label{pds_simul}
\end{figure}

However, some trend deviation or power excess is
noticeable in two cases.  For day 39, we fitted these PDSs with
a broken power law with an additional Lorentzian (the blue line in Fig.~\ref{pds}). This additional
component suggests the presence of a QPO, and the fits show 
   marginally improved $\chi^2_{\rm red}$ values. The improvement decreases 
 with increasing $n_{\rm div}$. Is
 this power fluctuation real, or is it the result of a random process? 
 In order to answer this question,
 we performed 10000 simulations of the light curves using both a simple 
broken power, and the method
 of \citet{timmer1995}. The best result 
 we obtained in simulating the PDS in two cases,
 with the day 38 {\sl Chandra} data are shown
 in Fig.~\ref{pds_simul}. The power fluctuations can be explained with a random process
 and we concluded that
 the variability in all three observations is dominated by a red noise.

  Finally, we also calculated the PDS of the optical light curve obtained with
 the {\sl XMM-Newton}
 OM. The light curve is presented in
 Fig. 11. We detected only red noise for frequencies lower than 0.001 Hz. Higher frequencies 
 are dominated by Poissonian noise.

\begin{figure}
\begin{center}
\resizebox{0.7\hsize}{!}{\includegraphics{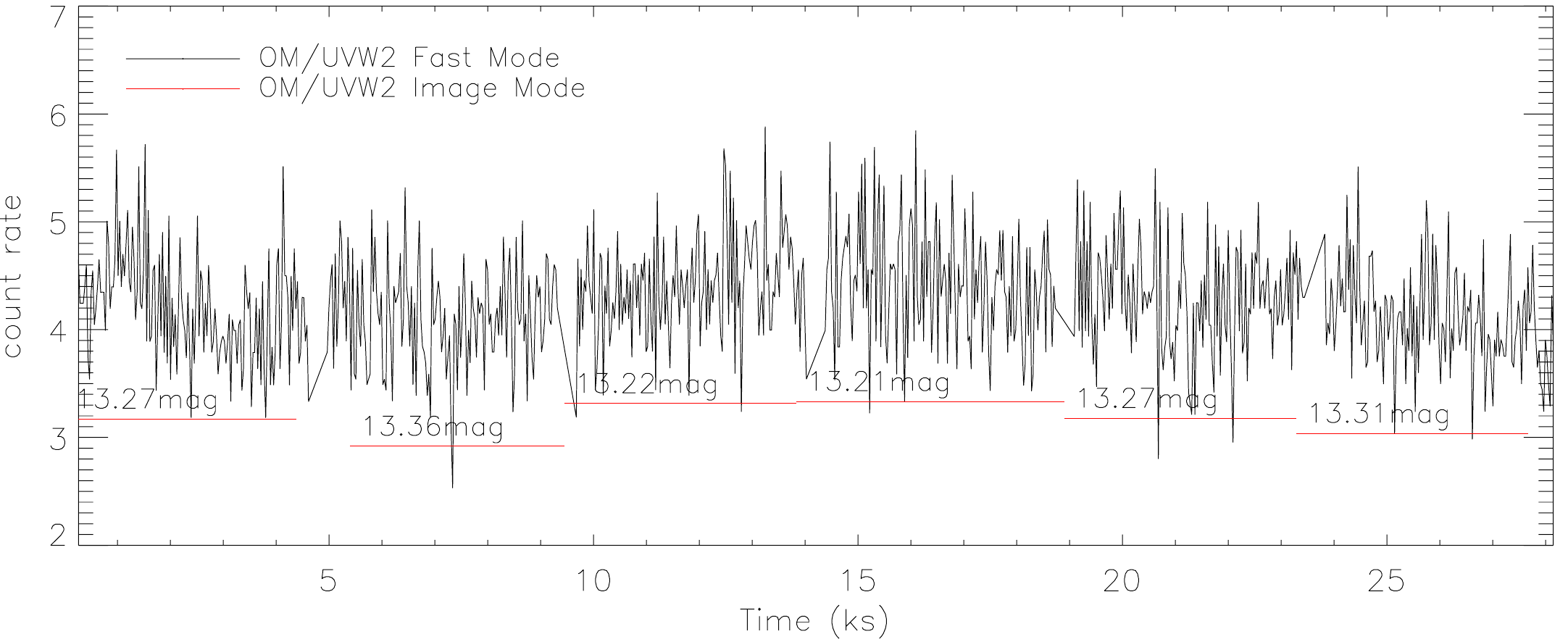}}
\end{center}
\caption{The UVW2 filter light curve observed with the XMM-Newton OM 
 on day 75, in fast mode and magnitudes derived
 in the given intervals in imaging mode, as described in the
 text.}
\label{omlc}
\end{figure}
\section{Discussion and Conclusions}

Using spectral fits to the X-ray data with 
 the TMAP model  of  static atmospheres of \citet{rauch2010},
we found that the WD in Nova SMC 2016 reached a peak T$_{\rm eff}$ between
 850,000 K and 900,000 K. In the nova models, such a temperature indicates a WD of $\simeq$1.25
 M$_\odot$, a conclusion also
 reached by \citet{aydi2018}. We also found
 absorption features of intermediate mass
 elements and magnesium
 that are not observed in V4743 Sgr, a nova observed at similar temperature
 in the SSS phase, that is thought to have occurred
 on a carbon-oxygen WD. This may imply that the WD here is
 of oxygen-neon, but more sophisticated modeling
 is still needed.  

 These observations highlight the potential of X-ray gratings to detect 
 very interesting physical characteristics of novae white dwarfs.
 The SSS source is the only phase in which we can observe nuclear burning
 very close to the stellar surface. Our
 spectra cannot be perfectly fitted with the published atmospheric grids,
 but the static models do match the absorption edges and the overall
 structure of the absorption features. 
 By experimenting with the WT model of  \citet{vanRossum2012}, 
 we found evidence that the residual wind from this nova
occurs in a different mode than the winds of massive O and B stars
 assumed as a template for this model,
 and/or it should involve a small amount of outflowing mass.
 In fact, assuming
 physics similar to the massive stars' winds, the model fails to account
 for the sharp absorption edges, deep absorption features and 
 drop in the flux of the softer continuum  that are observed  in this nova. 

 The fit with the TMAP atmospheric
 model fine-tuned for the abundances of V4743 Sgr \citep{rauch2010},
reproduces the deep nitrogen absorption features,
 for day 39 with [N/N$_{\rm odot}$]=1.803 and [C/C$_{\rm odot}$]=-1.513. 
 This implies a very large N/C ratio, as
 expected for the ashes of CNO burning, 
 and likely indicates that the burning
 is not occurring in freshly accreted material  after the thermonuclear
 runaway, which would dilute the abundances, and,  on the other
 hand,  probably also not in core-dredged matter (which would have a different
 composition, and be richer in carbon). A highly enhanced
  nitrogen to carbon (N/C) ratio, as we found, is typical
 only of the atmosphere on an envelope that has  been accreted and,
  in large part, already burned.
   Thus, this nova is probably burning in a retained a portion of the
 previously accreted and not ejected
 envelope, leaving the possibility open that the WD may be increasing in mass over
 its secular evolution. The TMAP fit for the spectra on days 75 and 88 indicates
 a decrease of the high N/C ratio, possibly indicating that accretion of
 material from the secondary has started again.

 We do not have
 any atmospheric model with abundances appropriate for a neon-oxygen WD yet, but
 because of  the spectral features of magnesium and of intermediate atomic mass
 elements (calcium, argon, and tentatively sulfur), and by comparison
 with other novae, we detect
 possible evidence of an enhancement of the above
 elements,  as expected for the peculiar nucleosynthesis on 
 oxygen-neon WDs  because of the neon-sodium and magnesium-aluminum
 cycles mentioned above. The abundances will be investigated with
 a fine-tuned model by some of us (Rauch et al. 2018, in preparation). 

 The X-ray grating spectra of novae yield copious physical information,
 but we are exploring ``virgin territory'', and are still developing
 the tools to achieve a thorough understanding. By identifying prominent spectral
 lines that indicate enhanced abundances and matching
 them with atmospheric models, we plan to obtain 
 the physical classification of the underlying WD.

 As noted by \citet{ness2011} for the spectra of three Galactic novae, we
 could not identify yet all the features in these intricate spectra.
 Some of these unidentified lines are common to different novae. It is 
 likely that we should ask the producers of atomic data to develop 
 the database to account for
 these lines; however, since we do not have a fine-tuned atmospheric model yet,
 we cannot completely rule out that 
 some unidentified features are produced by photoionization
 or even collisional ionization far from the WD,
 namely in the ejecta, at different velocity than the system
 of atmospheric lines.

 The X-ray gratings' spectra of N SMC 2016 also show the potential of the
 grating spectra of luminous novae in the Magellanic Clouds to yield
the chemical composition of the ISM along their lines of sight,
using the novae as lamps.

 Moreover, the long exposures needed to obtain
 grating spectra have yielded new information on the short term light
 curves of many novae, on time scales of hours. In this nova
 we detected large irregular variability and red noise,
 but no clear periodicities. Quite surprisingly, we did not find evidence
 that the flux variability was associated with variable absorption. We do
 not think that this necessarily means that the variability is intrinsic
 to the WD atmosphere;  we hypothesize that intermittent
bursts of mass ejection 
 caused the variability, with ejection of material that was completely
 optically thick to the SSS radiation, but was not ejected in a spherically 
 symmetric manner, hiding a portion of the WD. This phenomenon 
 would explain the non-absorption dependent variability, and if this
 is the correct explanation for it, the outflow of matter from the nova
 must be variable or intermittent on a timescale of hours. 
 So far, intermittent mass ejection, although on time scales
 of weeks, has only been invoked to explain the
 light curve of T Pyx \citep[][]{Chomiuk2014}. However,
 we note that also the complete disappearance of the WD, 
 for hours at the end of one
 X-ray observation of V4743 Sgr \citep[][]{ness2003}, seems to be best
 explained with a new ``burst'' of ejected mass that was optically thick
 to the supersoft X-rays. 

 Since the nova has returned to a quiescent status and the nova shell is expanding
 away from the WD, while accretion should
 have resumed, in the near future we hope to obtain an optical spectrum of the central accreting
 source and measure whether systemic velocity is indeed compatible with SMC membership. This will confirm that N SMC 2016 exceeded Eddington luminosity 
 for  more than 3 months, probably close to half a year, and
 belongs to  the rare group of superluminous novae.   
\facilities{Chandra, XMM-Newton}
\software{XSPEC
 \citep[v12.6.0][]{Arnaud1996}, CIAO \citep[v4.9;][]{Fruscione2006}, XMM-SAS v16.1.0. }
\bibliographystyle{apj}
\bibliography{biblio}
\end{document}